\documentclass[twocolumn,trackchanges]{aastex701}

\usepackage{multirow}
\usepackage{placeins} 
\usepackage{xcolor}
\usepackage{bm}
\usepackage{amsmath}

\AtBeginDocument{\hypersetup{allcolors=black}}

\begin{document}

\title{Planetary Habitable Zones around Magnetized White Dwarfs}
\correspondingauthor{Cong Yu}
\email{yucong@mail.sysu.edu.cn}

\author[orcid=0009-0002-7934-8562, gname=Xueyuan, sname=Wang]{Xueyuan Wang}
\affiliation{School of Physics and Astronomy, Sun Yat-sen University, Zhuhai, China}
\affiliation{International Centre of Supernovae (ICESUN), Yunnan Key Laboratory, Kunming 650216, China}
\email{Xueyuanwang4@gmail.com}  

\author[orcid=0000-0003-0454-7890, gname=Cong, sname=Yu]{Cong Yu} 
\affiliation{School of Physics and Astronomy, Sun Yat-sen University, Zhuhai, China}
\affiliation{International Centre of Supernovae (ICESUN), Yunnan Key Laboratory, Kunming 650216, China}
\affiliation{CSST Science Center for the Guangdong-Hong Kong-Macau Greater Bay Area, Zhuhai 519082, China}
\affiliation{State Key Laboratory of Lunar and Planetary Sciences, Macau University of Science and Technology, Macau, China}
\email[]{}

\author[orcid=0000-0003-0378-2023, gname=Bo, sname=Ma]{Bo Ma} 
\affiliation{School of Physics and Astronomy, Sun Yat-sen University, Zhuhai, China}
\affiliation{CSST Science Center for the Guangdong-Hong Kong-Macau Greater Bay Area, Zhuhai 519082, China}
\email[]{}

\author[orcid=0000-0003-0707-3213, gname=Di-Chang, sname=Chen]{Di-Chang Chen} 
\affiliation{School of Physics and Astronomy, Sun Yat-sen University, Zhuhai, China}
\affiliation{CSST Science Center for the Guangdong-Hong Kong-Macau Greater Bay Area, Zhuhai 519082, China}
\email[]{}

\author[0000-0002-8869-6510]{Jianheng Guo}
\affiliation{Yunnan Observatories,
Chinese Academy of Sciences, P.O. Box 110, Kunming 650011, China}
\affiliation{University of CAS, Beijing, China}
\affiliation{Key Laboratory for the Structure and Evolution of Celestial Objects, CAS, Kunming 650011, China.}
\email{}

\begin{abstract}
White dwarfs are known to host remnant planetary material, indicating that rocky objects can survive post-main-sequence evolution and migrate into very close orbits. However, the decline of white dwarf luminosity makes it difficult for a planet to remain habitable for long when relying only on the white dwarf's irradiative heating. While several non-radiative energy sources have been proposed, the role of white dwarf magnetism in planetary energy budgets remains largely unexplored. We examine whether magnetic unipolar induction could provide sustained heating for close-in rocky planets. By coupling white dwarf cooling tracks with an orbit-dependent induction model, we evaluate the contribution of ohmic dissipation to planetary surface energy budgets and estimate the upper limit with efficient heat deposition. It is possible for the induction heating to supply a nearly steady surface energy flux for certain orbital distances and stellar spins, partially offsetting the decline in the white dwarf irradiation at late times. This leads to a persistent outer temperate region and, in some cases, a narrow inner zone that emerges after substantial cooling, identifying magnetic induction as a viable long-lived supplement to radiative heating.
\end{abstract}


\section{Introduction}

White dwarfs (WD) represent the most common final stage of stellar evolution \citep{Fontaine_2001}. After exhausting nuclear fuel and shedding their envelopes, they cool for billions of years while steadily fading in luminosity \citep{Althaus_2010}. Their compact sizes and long cooling ages allow white dwarfs to serve as powerful laboratories for exploring the long–term dynamical fate of planetary systems \citep{Veras_2016}. Observations show that planets and smaller bodies frequently survive during the white dwarf phase, as evidenced by metal–polluted atmospheres, debris disks, disintegrating planetesimals, and at least one intact giant planet transiting a white dwarf \citep{Koester_2014,Vanderburg_2015,Vanderburg_2020,ChenDC19}. The short diffusion timescales of metals further imply continuous inward delivery of planetary material
\citep{Koester_2009,Zuckerman_2010}, with no observational evidence for a strong suppression of accretion
at late cooling ages \citep{Blouin_2021}.

Post main sequence mass loss can disrupt planetary systems and scatter planets, moons, and asteroids onto close orbits around white dwarfs \citep{Mustill_2012,Veras_2016}, naturally placing rocky bodies at small orbital distances where their surface environments may be strongly modified by nonradiative processes \citep{Veras_2014a,Veras_2014b}. Because white dwarfs are faint, their radiative habitable zones lie close to the star, typically at distances of order $10^{-2} \mathrm{AU}$ \citep{Agol_2011}, and migrate inward rapidly as the star cools \citep{Althaus_2010}. The cooling history therefore plays a central role in determining whether close–in rocky planets can maintain surface liquid water \citep{Barnes_2013,Kozakis_2018}. At the same time, the comparable radii of Earth-sized planets and white dwarfs enable exceptionally favorable transit spectroscopy, allowing atmospheric biosignatures to be detected within only a few hours of \textit{JWST} observations \citep{Loeb_2013}.

Recent work has expanded the energy budget of planets orbiting white dwarfs beyond stellar irradiation. Tidal heating can modify celestial bodies surface temperatures and enable transient habitability outside the classical radiative habitable zone \citep{Heller_2013}, while also driving inward orbital migration that allows planets to pass through regions where tidal dissipation may help sustain liquid water \citep{Becker_2023}. In addition to such planet-level energy sources, white dwarfs themselves may undergo nonstandard thermal evolution. Advances in stellar evolution modeling show that processes such as $^{22}\mathrm{Ne}$ diffusion and distillation can substantially delay the cooling of massive white dwarfs, producing luminosity histories that depart from standard cooling tracks \citep{Camisassa_2016,Salaris_2021}. Such delayed cooling can significantly alter the extent and duration of radiatively permitted surface conditions for close-in rocky planets, shifting or extending white dwarf habitable zones \citep{Vanderburg_2025}. Together, these studies demonstrate that additional internal or external energy sources can strongly influence habitability around white dwarfs.

By contrast, the role of white dwarf magnetism in planetary energy budgets has not yet been incorporated into habitability studies. Observational surveys indicate that about $10$--$20\%$ of white dwarfs host detectable magnetic fields \citep{Ferrario_2015,Ferrario_2020,Bagnulo_2021}, implying that magnetic star--planet interactions are likely common. These fields are both strong and exceptionally long-lived, with surface strengths of $10^{5}$--$10^{9}\,\mathrm{G}$ \citep{Ferrario_2020,Bagnulo_2021} and large-scale configurations that decay only on Ohmic timescales of $\gtrsim10^{9}$--$10^{12}\,\mathrm{yr}$ \citep{Cumming_2002,Ferrario_2015}, far exceeding typical white dwarf cooling ages. Therefore, the magnetosphere is effectively steady over planetary evolution timescales. Planets migrating into close orbits will inevitably cross strong magnetic fields.

A conducting body moving through such a field develops an induced electromotive force, drives field-aligned currents, and dissipates energy within the planet or at the stellar magnetic footpoints \citep{Goldreich_1969}. Originally developed to describe the Jupiter--Io interaction, the unipolar--inductor mechanism has since been extended to star--planet systems, with observational signatures of sub-Alfv\'enic magnetic interactions inferred from low-frequency radio emission \citep{Vedantham_2020} and persistent H$\alpha$ emission in magnetic white dwarfs such as GD~356 \citep{Li_1998}. Theoretical studies of magnetically interacting close binaries and star--planet systems further demonstrate how orbital motion in strong magnetic fields can generate substantial resistive dissipation \citep{Laine_2008,Lai_2012,Laine_2012}. High-field white dwarf magnetospheres can therefore supply large electromagnetic energy fluxes \citep{Ferrario_2015}, a fraction of which, if coupled through a unipolar--inductor circuit, may provide significant non-radiative heating to close-in rocky planets. Electromagnetic induction in solid bodies orbiting magnetic white dwarfs has been shown to generate substantial ohmic dissipation and to drive orbital energy loss, particularly for small rocky and metallic objects on close-in or eccentric orbits \citep{Bromley_2019}. Although these studies focused on internal heating and orbital evolution rather than surface habitability, they demonstrate that magnetic induction can act as an efficient and long-lived non-radiative energy source in white dwarf systems.

Related work has shown that time-varying magnetic fields can induce strong ohmic dissipation within the interiors of Moon-sized rocky bodies, leading to partial melting and volcanism \citep{Kislyakova_2023}. However, these studies focused primarily on internal heating and geological consequences, and did not assess the impact of magnetic star--planet interactions on planetary surface energy balance, the evolution of surface temperatures along white dwarf cooling tracks, or the conditions required to sustain surface liquid water.

The stellar rotation rate is a key control on induction heating. In the unipolar--inductor framework, the induced electromotive force scales with the relative angular velocity between the star and the planet \citep{Lai_2012}, such that rapidly rotating white dwarfs or planets on slower orbits generate strong induction voltages and Poynting fluxes, while nearly synchronous systems produce much weaker currents \citep{Laine_2008,Laine_2012}. Observed white dwarf spin periods span hours to days \citep{Hermes_2017}, allowing induction heating to operate across a wide range of orbital configurations.

The efficiency of induction heating further depends on a planet’s internal electrical conductivity. Metal-rich planets with molten or partially molten iron cores can sustain large currents and experience strong ohmic dissipation, whereas dry, silicate-dominated interiors are far more resistive, with water-rich or ocean-bearing worlds occupying an intermediate regime. This sensitivity is illustrated by the contrast between Io and Europa: despite residing in a similar magnetic environment, Io’s conductive, partially molten interior supports intense field-aligned currents and extreme volcanism \citep{Kivelson_2001}, while Europa’s layered ice--ocean--rock structure leads to much weaker coupling \citep{Saur_2010}. Magnetometer measurements from \textit{Galileo} confirm Europa’s induced magnetic response and imply a conductive subsurface ocean \citep{Khurana_1998,Zimmer_2000}. Together, these examples demonstrate that induction heating rates can vary by orders of magnitude for planets exposed to the same stellar magnetic field.

In this work, we couple white dwarf cooling tracks with an orbit-based model of unipolar induction heating to explore the thermal evolution of close-in rocky planets. By surveying a broad range of orbital distances and stellar spin periods, we examine how magnetic induction alters the conventional picture of white dwarf habitability and assess whether ohmic dissipation can sustain surface liquid water at late cooling ages.

\begin{figure*}[!t]
    \centering
    \includegraphics[width=\textwidth]{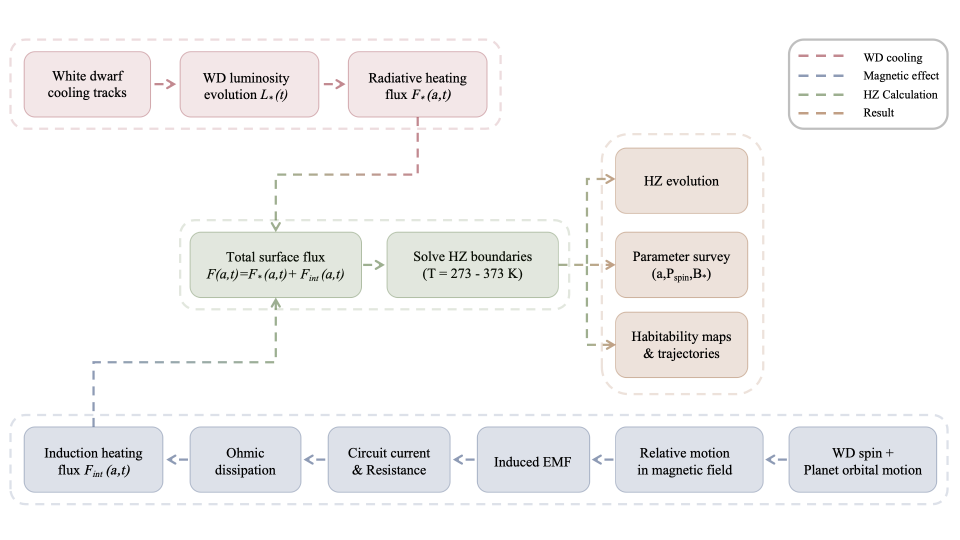}
    \caption{
    Schematic overview of the computational framework used in this work. 
    The upper branch describes the radiative heating channel based on white dwarf cooling tracks, while the lower branch represents the magnetic unipolar-induction process that generates induction heating through relative spin-orbit motion. 
    These two energy contributions are combined to compute the total planetary surface flux and determine the habitable-zone evolution. 
    The resulting model is then applied to parameter surveys and the construction of habitability maps and evolutionary trajectories.
    }
    \label{fig:workflow}
\end{figure*}

Figure~\ref{fig:workflow} summarizes the computational framework adopted in this work, including the coupled treatment of white dwarf irradiation, magnetic induction heating, and habitable-zone evolution.

This paper is organized as follows. Section~\ref{sec:energy} describes the two energy sources considered in this work, including stellar irradiation along white dwarf cooling tracks and heating due to unipolar magnetic induction. Section~\ref{sec:hz_combined} combines irradiation and induction heating to compute the resulting habitable zone. Sections~\ref{sec:results} and~\ref{sec:discussion} present the main results and discuss limitations, observational implications, and the broader role of magnetic induction in shaping white dwarf habitability.

\section{Energy sources}
\label{sec:energy}

\subsection{WD irradiation}
\label{sec:wd_irradiation}

The radiative environment of a close-in white dwarf planet is set by the stellar luminosity, which declines monotonically as the white dwarf cools in the absence of nuclear energy generation. Following the equilibrium surface-temperature formulation of \citet{Becker_2023}, we write
\begin{equation}
\label{eq:WD_term}
    \sigma T^4 = \frac{(1-A)L_\star}{16\pi a^2},
\end{equation}
and adopt an Earth-like Bond albedo of $A=0.3$. As the luminosity decreases with cooling age, the habitable zone (HZ) maintained by the WD irradiation heating migrates steadily inward, making the stellar cooling history a central ingredient in any assessment of long-term habitability.

\begin{figure}[!t]
    \centering
    \includegraphics[width=\linewidth]{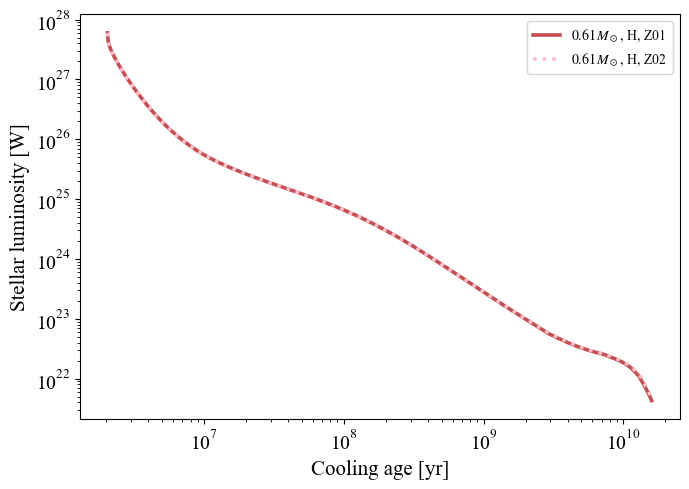}
    \caption{
    Luminosity evolution of a $0.61\,M_\odot$ carbon--oxygen white dwarf with a hydrogen atmosphere for two progenitor metallicities, $Z=0.01$ and $Z=0.02$, based on the cooling tracks of \citet{Salaris_2021}. }
    \label{fig:wd_cooling_luminosity}
\end{figure}

We model the luminosity evolution using the carbon--oxygen core white dwarf cooling tracks of \citet{Salaris_2021}. The cooling sequence adopted here extends to a total age of $t_{\rm Gyr}=15.95$, providing full coverage from early hot stages to late, low-luminosity evolution. Because the radiative habitable-zone distance scales as $a \propto L_\star^{1/2}$, even modest differences in the cooling history can lead to appreciable shifts in the location and duration of radiatively allowed surface conditions.

Figure~\ref{fig:wd_cooling_luminosity} shows the luminosity evolution of a $0.61\,M_\odot$ hydrogen-atmosphere white dwarf for two progenitor metallicities, $Z=0.01$ and $Z=0.02$.

Over the cooling ages relevant to this work, the two tracks are closely similar. This near-indistinguishability is established only for the two directly compared progenitor metallicities, $Z=0.01$ and $Z=0.02$. We therefore restrict our conclusion to the interval $Z\in[0.01,0.02]$ and do not extrapolate it to lower or higher metallicities. We adopt the $0.61\,M_\odot$ cooling sequence as a representative radiative input for the induction calculations.

\subsection{Unipolar induction}
\label{sec:unipolar_induction}
Electromagnetic induction provides an additional, non-radiative energy source for close-in white dwarf planets. When a conducting planet orbits within the magnetosphere of a magnetic white dwarf, relative motion between the orbital velocity and the stellar magnetic field induces an electromotive force (EMF), driving a current loop that connects the planet and the star through Alfvén wings \citep{Drell_1965, Neubauer_1998}. This process follows the classical ``DC circuit'' or unipolar--inductor framework, originally developed to describe the Jupiter--Io interaction and later extended to magnetized star--planet and compact binary systems \citep{Goldreich_1969, Li_1998, Laine_2012, Willes_2004, Bagenal_2020}. The resulting ohmic dissipation can heat the planetary  atmospheric boundary layer on long timescales. Below we summarize the ingredients of the model as implemented in our calculations.

\subsubsection{Resistance}

The total resistance used in our closed-DC calculation is
\begin{equation}
    \mathcal{R}_{\rm tot}
    = \mathcal{R}_{\rm p}
    + \mathcal{R}_{\star}
    + \mathcal{R}_A ,
\end{equation}
where $\mathcal{R}_{\rm p}$ is the planetary resistance, $\mathcal{R}_{\star}$ is the effective resistance of the stellar magnetic footpoints, and $\mathcal{R}_A$ represents the effective field-aligned resistance of the magnetospheric flux tube
\citep{Li_1998,Laine_2012,Lai_2012}.

Following \citet{Laine_2012}, the flux-tube resistance can be estimated as
\begin{equation}
\mathcal{R}_A
\sim
\frac{a}{\pi R_{\rm p}^{2}\sigma_{\parallel}f_{\rm tube}},
\end{equation}
where $f_{\rm tube}\leq1$ parameterizes the thickness of the current-carrying region.
For $a=0.02\,{\rm AU}$, $R_{\rm p}=R_{\oplus}$, and
$\sigma_{\parallel}=2\times10^{3}\,{\rm S\,m^{-1}}$,
\begin{equation}
\mathcal{R}_A
\simeq
1.2\times10^{-8}f_{\rm tube}^{-1}\ {\rm \Omega}.
\end{equation}
This is negligible compared with
$\mathcal{R}_{\rm p}+\mathcal{R}_{\star}=1.8\,\Omega$
unless the current is confined to an extremely thin layer.
We therefore use
$\mathcal{R}_{\rm tot}\simeq
\mathcal{R}_{\rm p}+\mathcal{R}_{\star}$
in the calculations below.

\textbf{Planetary resistance.}
Following the standard treatment of electrodynamic coupling in satellite--planet and star--planet systems, we model the planet as a spherically stratified electrical conductor composed of a metallic core, a silicate mantle and crust, an optional ocean layer, and a tenuous atmosphere \citep{Neubauer_1998,Saur_2004}. This layered approach captures the strong radial variation in electrical conductivity expected for terrestrial planets and has been widely adopted in studies of sub--Alfv\'enic magnetic interactions.

Each layer is assigned a representative electrical conductivity $\sigma_i$. For the metallic core, we adopt
$\sigma \sim 10^{5}\,\mathrm{S\,m^{-1}}$. Dry silicate mantle and crustal material are assigned $\sigma \sim 10^{-4}\,\mathrm{S\,m^{-1}}$ as a fiducial poorly conducting rocky value; the strong temperature dependence of rocky-planet conductivity is discussed by \citet{Laine_2008}. Where present, an ocean layer is assigned an intermediate conductivity of order
$\sigma \sim 1$--$5\,\mathrm{S\,m^{-1}}$, consistent with saline water and supported by magnetometer-based inferences of subsurface oceans in icy satellites \citep{Zimmer_2000,Neubauer_1998}. 

The atmospheric boundary layer is assigned
a low effective radial conductivity, $\sigma_{\rm atm}=10^{-8}\,\mathrm{S\,m^{-1}}$.
We adopt an effective atmospheric thickness of $\Delta r_{\rm atm}=0.15R_{\rm p}$ (approximately $960\,\mathrm{km}$ for an Earth-sized planet), corresponding to a total current-path length of $\ell_{\rm atm}=2\Delta r_{\rm atm}=0.30R_{\rm p}$.
Using the effective cross-sectional area $\pi R_{\rm p}^{2}$, this gives an atmospheric resistance of $\mathcal{R}_{\rm atm}\simeq1.50\,\Omega$.

The adopted solid-layer thicknesses are broadly Earth-like, with the metallic core extending to $\sim0.6\,R_{\rm p}$ and the silicate mantle occupying most of the remaining solid volume. These values and the effective atmospheric path are not intended to define a unique planet, but provide an explicit reference structure for an order-of-magnitude calculation.

In the fiducial model, the mantle and ocean layer thicknesses are $\Delta r_{\rm mantle}=0.399R_{\rm p}$ and $\Delta r_{\rm ocean}=0.001R_{\rm p}$, respectively, with an ocean conductivity of $\sigma_{\rm ocean}=2\,\mathrm{S\,m^{-1}}$.

The adopted atmospheric thickness represents an effective resistive path rather than a self-consistently modeled atmospheric structure.
Although planetary upper atmospheres can extend
to large altitudes, maintaining such a low conductivity across a $\sim1000\,\mathrm{km}$ layer is uncertain, since ionization and magnetic-field geometry can produce strongly altitude-dependent and anisotropic conductivities.
We therefore regard the atmospheric thickness and conductivity as effective model parameters rather than established properties of terrestrial planets around white dwarfs.

Approximating the dominant current path as extending across the planetary diameter, the electrical resistance of an individual layer is parameterized as
\begin{equation}
\mathcal{R}_i \simeq
\frac{\ell_i}{\sigma_i \,\pi R_{\rm p}^{2}},
\end{equation}
where $\ell_i=2\Delta r_i$ is the total path length through the layer on the two sides of the planet and $\pi R_{\rm p}^{2}$ is the adopted effective cross-sectional area \citep{Laine_2012}. The total planetary resistance is then
\begin{equation}
    \mathcal{R}_{\rm p} = \mathcal{R}_{\rm atm} + \mathcal{R}_{\rm ocean} + \mathcal{R}_{\rm mantle} + \mathcal{R}_{\rm core}.
\end{equation}

We have listed the results used in the calculations in Table \ref{tab:phys_limits_example}.

\textbf{Stellar footpoint resistance.}
The resistance associated with current closure at the white dwarf magnetic footpoints is highly uncertain, as it depends on poorly constrained properties of the surface layers, including their conductivity profiles, ionization state, and magnetic topology\citep{Ferrario_2015}. In studies of compact-object electrodynamics, this contribution is commonly parameterized through an effective footpoint resistance that captures the net ability of the stellar surface to close the unipolar--induction circuit \citep{Li_1998, Lai_2012}.

Instead of adopting a detailed layered model for the white dwarf atmosphere, which would introduce several poorly constrained parameters, we use $\mathcal{R}_{\star,0}=0.3\,\Omega$ as a fiducial effective footpoint resistance and explicitly test its cooling dependence below.

As the white dwarf cools from $T_{\rm eff}\sim10^4\,\mathrm{K}$ to $\sim3000\,\mathrm{K}$, recombination reduces the atmospheric electron fraction \citep{Rohrmann_2001} and may increase the effective footpoint resistance. Because the mapping from ionization fraction to the global circuit load remains poorly constrained, we use the following parameterized sensitivity prescription rather than claiming a microphysical prediction:
\begin{equation}
\mathcal{R}_{\star}(T_{\rm eff})=\mathcal{R}_{\star,0}
\left(\frac{T_{\rm eff}}{10^4\,{\rm K}}\right)^{-\beta},
\qquad
\beta=\frac{\ln f_{\mathcal R}}{\ln(10^4/3000)},
\end{equation}
where $\mathcal{R}_{\star,0}=0.3\,\Omega$ is the value at $10^4\,\mathrm{K}$ and $f_{\mathcal R}\equiv\mathcal{R}_{\star}(3000\,{\rm K})/\mathcal{R}_{\star}(10^4\,{\rm K})$ specifies the endpoint growth. The resistance is held at the corresponding endpoint outside the interval $3000\,\mathrm{K}\leq T_{\rm eff}\leq10^4\,\mathrm{K}$. We test $f_{\mathcal R}=$1, 3, 10, and 100. The constant-$\mathcal{R}_{\star}$ baseline is an optimistic limiting case, whereas the cooling-dependent models test the longevity of induction heating (Section~\ref{sec:rstar_sensitivity}).

\textbf{Alfv\'en communication.} Alfv\'en characteristics communicate the disturbance between the planet and the white dwarf.

Following \citet{Neubauer_1980}, the Alfv\'enic conductance in the strongly sub-Alfv\'enic limit is
\begin{equation}
\Sigma_A \simeq \frac{1}{\mu_0 v_A}.
\end{equation}
For the fiducial value $v_A=2.68\times10^{8}\,{\rm m\,s^{-1}}$,
this gives $\Sigma_A\simeq3.0\times10^{-3}\,{\rm S}$.
This conductance characterizes Alfv\'enic communication and is distinct from the dissipative flux-tube resistance $\mathcal{R}_A$ defined above.
The conditions for establishing the closed DC circuit are tested in Section~\ref{sec:physical_limits}.

\subsubsection{Induced EMF}

The orbital motion of the planet through the rotating dipolar magnetic field of the white dwarf generates a motional electromotive force (EMF). We adopt a static dipole field, such that the magnetic strength at orbital distance $a$ is
\begin{equation}
    B(a) = B_0 \left( \frac{R_\star}{a} \right)^3 .
\end{equation}
The relative angular velocity between the planet and the stellar surface is defined as
$\Delta \Omega = \big|\Omega_{\rm orb} - \Omega_{\rm spin}\big|$,
which gives a relative linear speed
$v_{\rm rel} = a\,\Delta\Omega$.
The induced EMF then takes the standard unipolar–inductor form,
\begin{equation}
    \mathcal{E}
    = 2 R_{\rm p} B(a) v_{\rm rel}
    = 2 R_{\rm p} B_0 \left(\frac{R_\star}{a}\right)^3 a
      \big|\Omega_{\rm orb}-\Omega_{\rm spin}\big|.
\end{equation}
This expression is equivalent to the usual $(\mathbf{v}\times\mathbf{B})\ell$ motional EMF used in unipolar--induction models \citep{Goldreich_1969, Li_1998, Lai_2012}. A key feature is that synchronous rotation ($\Omega_{\rm orb}=\Omega_{\rm spin}$) yields $v_{\rm rel}=0$ and therefore $\mathcal{E}=0$, so sustained induction heating requires non-synchronous rotation, as expected for most white dwarf--planet systems.

\subsubsection{Heating power}

Given the EMF and the total circuit resistance
$\mathcal{R}_{\rm tot} = \mathcal{R}_{\rm p} + \mathcal{R}_{\star}$

(the flux-tube resistance $\mathcal{R}_A$ is neglected as justified above),
the circuit current is
\begin{equation}
    I = \frac{\mathcal{E}}{\mathcal{R}_{\rm tot}} .
\end{equation}

The total electromagnetic power extracted from the circuit follows as
\begin{equation}
    P_{\rm tot}
    = \mathcal{E}\, I
    = \frac{\mathcal{E}^2}{\mathcal{R}_{\rm tot}} .
\end{equation}
as in standard unipolar--inductor and DC-circuit models
\citep{Goldreich_1969, Li_1998, Lai_2012}.

Only a fraction of this power is dissipated  within the planet
The planetary ohmic power is
\begin{equation}
    P_{\rm p} = I^2\,\mathcal{R}_{\rm p},
\end{equation}
corresponding to an effective surface heating flux
\begin{equation}
    F_{\rm p} = \frac{P_{\rm p}}{4\pi R_{\rm p}^2},
\end{equation}
The deposited flux used in the thermal calculation is defined below.

Since the same current $I$ flows through each effective resistive element in our framework, the layer-by-layer dissipation is given by $P_i=I^2\mathcal{R}_i$. As established previously, the atmospheric resistance is orders of magnitude larger than that of the solid interior ($\mathcal{R}_{\rm atm} \gg \mathcal{R}_{\rm solid}$).

As our model does not explicitly resolve how much of the atmospheric boundary-layer ohmic power contributes to the surface energy budget, we introduce an energy-coupling factor $0\leq f_{\rm dep}\leq1$, where
\[
F_{\rm eff}=f_{\rm dep}\,F_{\rm p} \ .
\]
Our habitable-zone maps use $f_{\rm dep}=1$, meaning that all planetary ohmic power is included in the surface heat budget. This choice sets an upper limit on temperature and habitable-zone changes.

Combining the above expressions shows that the planetary heating power scales as
\begin{equation}
    P_{\rm p} \propto
    \frac{B(a)^2\,R_{\rm p}^2\,a^2\,
          \big|\Omega_{\rm orb}-\Omega_{\rm spin}\big|^2}
         {(\mathcal{R}_{\rm p}+\mathcal{R}_{\star})^2}
    \,\mathcal{R}_{\rm p}.
\end{equation}
consistent with the scaling relations derived in previous treatments of magnetic star--planet induction \citep{Li_1998, Lai_2012}. 

The induced power therefore increases with the square of the local magnetic-field strength and with the degree of non-synchronous rotation, since both enhance the motional EMF. Larger planets intercept more magnetic flux and thus generate stronger induction voltages. The relative magnitudes of $\mathcal{R}_{\rm p}$ and $\mathcal{R}_{\star}$ determine where the dissipation occurs: when the planetary resistance dominates, most of the ohmic power is deposited on the planetary side of the circuit, predominantly in the atmospheric boundary layer in our model, whereas a larger stellar footpoint resistance would shift a larger fraction of the dissipation to the white dwarf footpoints \citep{Li_1998, Lai_2012}.

The distribution of electromagnetic dissipation within the circuit depends on the resistive properties of both the planet and the stellar magnetic footpoints. The total resistance controls the circuit current, while the partitioning of ohmic dissipation is set by the individual resistive elements, with energy deposited wherever current closure occurs within the system. 

In the following sections, we focus on the planetary contribution to the dissipation. Its application as a surface heat source adopts the upper-limit choice $f_{\rm dep}=1$ defined above.  We then assess its impact on the surface thermal evolution of close-in white dwarf planets.

\subsection{Physical Limits}
\label{sec:physical_limits}

The unipolar--induction model is subject to several physical validity conditions that restrict the range of orbital separations and circuit parameters for which a steady electromagnetic coupling can exist. In particular, the formation of a coherent DC circuit is not automatic, but requires that magnetic stresses and electrical currents be communicated efficiently between the planet and the host white dwarf without triggering magnetic instabilities or circuit breakdown. Here we summarize the key criteria adopted in this work and their physical motivation.

\subsubsection{ DC circuit}
\label{DC circuit}
A steady DC circuit first requires a sub-Alfv\'enic interaction.
This allows magnetic stresses to propagate between the planet and the star
along Alfv\'en characteristics
\citep{Goldreich_1969,Drell_1965,Neubauer_1998,Laine_2008,Lai_2012,Saur_2013}.
In the dilute magnetospheres expected around isolated white dwarfs,
the combination of strong magnetic fields and low plasma densities can produce
large Alfv\'en speeds, allowing magnetic disturbances to communicate between
the planet and the stellar surface. This sub-Alfv\'enic communication is a
necessary condition for establishing a steady unipolar-inductor or DC circuit
\citep{Laine_2008,Li_1998,Lai_2012,Saur_2013}.

We evaluate the Alfv\'en speeds as

\begin{equation}
\bm{
v_A=\frac{c\,v_{A,0}}{\sqrt{c^2+v_{A,0}^2}},
}
\end{equation}
where $v_{A,0} = B/\sqrt{\mu_0 \rho}$. The magnetic Mach number 
\begin{equation}
\bm{
    M_A = \frac{v_{\rm rel}}{v_A} < 1,
}
\end{equation}
where $v_{\rm rel}$ is the relative velocity between the planet and the
rotating magnetic field and $v_A$ is the Alfv\'en speed of the ambient plasma.

Second, we require the Alfv\'en round-trip time to be shorter than the time for the planet to move by one diameter along its orbit.
\citep{Goldreich_1969,Neubauer_1998,Li_1998,Lai_2012}:

\begin{equation}
\label{eq:dc_communication}
\bm{
    \tau_A = \frac{2a}{v_A}
    <
    \tau_{\rm cross}
    =
    \frac{2R_{\rm p}}{v_{\rm rel}},
}
\end{equation}
where $\tau_A$ is the Alfv\'en round-trip communication time,
$\tau_{\rm cross}$ is the magnetic crossing time for one planetary diameter, and
$v_{\rm rel}=a|\Omega_{\rm orb}-\Omega_{\rm spin}|$
is the relative spin--orbit velocity. This constraint should be satisfied for the adopted DC approximation.

The communication condition also imposes an upper limit on the ambient plasma number density. Assuming a cold, proton-dominated plasma with mass density $\rho=nm_{\rm p}$, we define the critical density $n_{\rm crit}$ by setting $\tau_A=\tau_{\rm cross}$. Substituting the relativistic Alfv\'en speed into this condition yields

\begin{equation}
\label{eq:ncrit_rel}
n_{\rm crit}
=
\frac{B(a)^2}{\mu_0 m_{\rm p}}
\left[
\frac{R_{\rm p}^2}{a^2v_{\rm rel}^2}
-
\frac{1}{c^2}
\right].
\end{equation}

Here $B(a)=B_\star(R_\star/a)^3$. The communication condition requires $n<n_{\rm crit}$, provided that $av_{\rm rel}/R_{\rm p}<c$. For the fiducial system, we obtain $n_{\rm crit}\simeq1.84\times10^3\,\mathrm{cm^{-3}}$, which exceeds the adopted plasma density of $n=100\,\mathrm{cm^{-3}}$. Thus, the communication condition remains satisfied.

\subsubsection{Critical resistance}

In addition to being sub–Alfv\'enic, the unipolar–induction circuit must remain magnetically stable. As shown by \citet{Lai_2012}, the total circuit resistance must exceed a critical value in order to prevent excessive magnetic twist from building up along the flux tube. This requirement can be written as
\begin{equation}
    \mathcal{R}_{\rm tot} \ge \mathcal{R}_{\rm crit},
\end{equation}
with the critical resistance given by
\begin{equation}
    \mathcal{R}_{\rm crit}
    \simeq 480
    \left( \frac{v_{\rm rel}}{c} \right)\,\Omega .
\end{equation}

If $\mathcal{R}_{\rm tot}$ falls below this threshold, the induced currents become strong enough to generate magnetic twist exceeding the steady-state limit, invalidating the DC-circuit approximation and leading instead to magnetic inflation or circuit breakdown.

For the fiducial configuration, the atmospheric layer dominates the planetary resistance, giving $\mathcal{R}_{\rm tot}\simeq1.80\,\Omega$, which exceeds
$\mathcal{R}_{\rm crit}\simeq0.433\,\Omega$. Without the atmospheric contribution, the total resistance would decrease to approximately $0.30\,\Omega$, falling below the critical threshold and invalidating the steady DC-circuit approximation for this configuration.
The critical-resistance condition requires
$\mathcal{R}_{\rm atm}\gtrsim0.133\,\Omega$
for the adopted fiducial parameters. 
Thus, a sufficiently resistive atmospheric
component is necessary for the present configuration, although the specific fiducial value of $1.50\,\Omega$ is not uniquely required.

\subsubsection{Poynting-flux limit}
A further constraint follows from energy conservation. Within the unipolar--inductor framework, the electromagnetic power extracted from orbital motion cannot exceed the rate at which magnetic stresses can be transported and dissipated along the connecting flux tube. For a dipolar stellar magnetic field, the maximum steady Poynting luminosity is given by \citet{Lai_2012}
\begin{equation}
    P_{\rm max}
    =
    \zeta_\phi\,
    \Delta\Omega\,
    \frac{\mu^2 R_{\rm p}^2}{2 a^5},
    \label{eq:poynting_limit}
\end{equation}
where $\mu = B_\star R_\star^3$ is the stellar magnetic dipole moment and $\zeta_\phi \sim 1$ parameterizes the azimuthal magnetic twist of the flux tube. Equation~\eqref{eq:poynting_limit} is expressed in Gaussian cgs units,
with $B_\star$ in gauss, $\mu$ in
$\mathrm{G\,cm^3}$, lengths in centimeters, and $P_{\rm max}$ in $\mathrm{erg\,s^{-1}}$. The resulting power is converted to watts for comparison with the total circuit power.

The total circuit power must therefore satisfy
\begin{equation}
    P_{\rm tot}(a) \le P_{\rm max}(a),
\end{equation}
which imposes a strict upper bound on the allowable ohmic dissipation in the unipolar--induction circuit.

Together with the  DC circuit requirement and the critical--resistance condition, this Poynting--flux limit delineates the physical regime in which the adopted DC approximation can be applied. Exceeding either the magnetic twist limit or the maximum extractable Poynting flux leads to magnetic inflation or circuit breakdown \citep{Lai_2012}, while super--Alfv\'enic motion or failure of the adopted round-trip contraint prevents rapid communication along the connecting field lines \citep{Goldreich_1969, Drell_1965, Neubauer_1998}. For the adopted fiducial system, $B_\star=100\,\mathrm{MG}$, $P_{\rm spin}=12\,\mathrm{h}$, $a=0.020\,\mathrm{AU}$, and $n=10^2\,\mathrm{cm^{-3}}$ satisfy these constraints. The present table verifies the condition only for the fiducial system.

\section{Habitability with Induction Heating}
\label{sec:hz_combined}


\begin{figure*}
    \centering
    \includegraphics[width=1\linewidth]{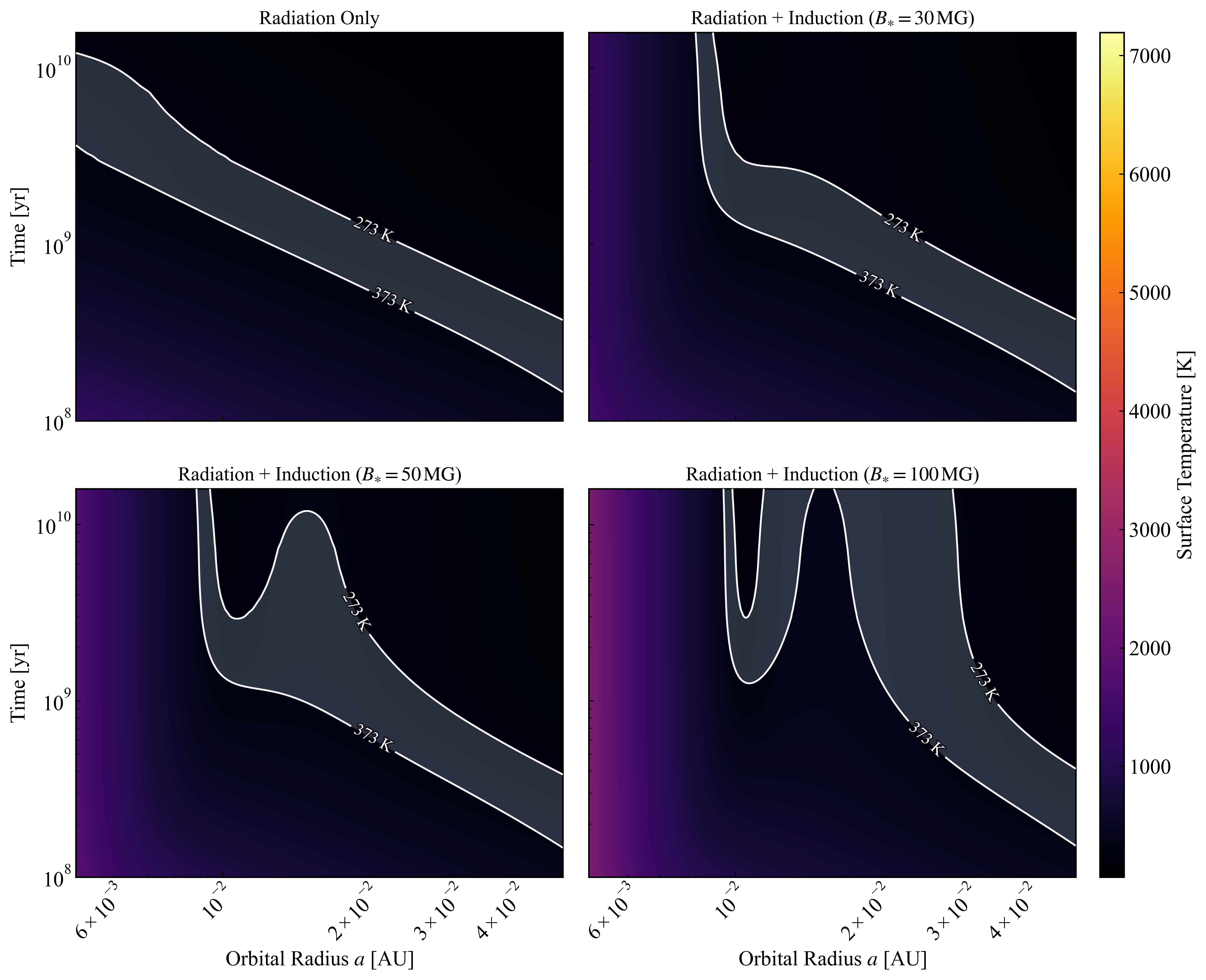}
    \caption{Surface-temperature evolution as a function of orbital radius and cooling time. The upper-left panel shows irradiation only; the other panels include the upper-limit planetary ohmic heating for $B_\star=$30, 50, and 100$\,\mathrm{MG}$, respectively, with $P_{\rm spin}=12\,\mathrm{h}$. The contours mark the 273--373 K interval used here as a temperate-flux proxy.}
    \label{fig:multi_field_hz}
\end{figure*}

To evaluate how unipolar induction affects the potential habitability of close-in white dwarf planets, we compute the time-dependent habitable zone (HZ) by combining stellar irradiation with the induction-driven heating derived in Section~\ref{sec:unipolar_induction}.  
At each cooling age $t$, the net surface flux at orbital radius $a$ is

\begin{equation}
\label{eq:Fnet}
F(a,t) = F_\star(a,t) + F_{\rm eff}(a)
\end{equation}

where the first term is the absorbed stellar radiation given by Eq.~(\ref{eq:WD_term}), and the second term is the deposited, surface-equivalent induction flux.
Using $\mathcal{E}=2R_{\rm p}B(a)a\Delta\Omega$ and  
$P_{\rm p}=I^2\mathcal{R}_{\rm p}=\mathcal{E}^2\mathcal{R}_{\rm p}/\mathcal{R}_{\rm tot}^2$,  
the   planetary ohmic heating flux may be written explicitly as
\begin{equation}
F_{\rm eff}(a)
= \frac{B(a)^2\, a^2\, \Delta\Omega^2}
       {\pi}\,
  \frac{f_{\rm dep}\mathcal{R}_{\rm p}}{\mathcal{R}_{\rm tot}^2}.
\end{equation}

In the baseline model, the stellar magnetic field, spin period, and circuit resistances are assumed to remain constant. Consequently, the induction heating flux
$F_{\rm eff}(a)$ is independent of the white dwarf cooling age at a fixed orbital radius. The time dependence of the total surface flux $F(a,t)$ therefore arises solely from the declining stellar irradiation. Unless orbital evolution is explicitly included, the results below refer to fixed orbital radii. The 273--373 K interval is a thermal proxy, not a prediction of surface liquid water; interpretation within the steady-DC model additionally requires the physical conditions in Section~\ref{sec:physical_limits}.

The inner and outer HZ radii are defined implicitly by  $\sigma T_{\rm in}^4 = F(a_{\rm in},t)$ and  $\sigma T_{\rm out}^4 = F(a_{\rm out},t)$,  with $T_{\rm  in}=373\,\mathrm{K}$ and $T_{\rm out}=273\,\mathrm{K}$.  Because the total surface flux depends nonlinearly on orbital radius through both the radial magnetic-field profile and the rotation-dependent induction heating term, the HZ boundaries cannot be expressed analytically and must be determined numerically. We therefore determine the HZ boundaries numerically by solving  $F(a,t)=\sigma T^{4}$ along the white dwarf cooling track.

To ensure dynamical viability, we further impose the tidal disruption constraint by excluding any orbital radius interior to the fluid Roche limit \citep{Chandrasekhar_1969}:
\begin{equation}
a_{\rm Roche} = 2.423\,R_\star
\left(\frac{\bar{\rho}_\star}{\bar{\rho}_{\rm p}}\right)^{1/3}.
\end{equation}

For the fiducial $0.61\,M_\odot$ white dwarf adopted in this work, with a stellar radius $R_\star \simeq 0.013\,R_\odot$ and mean density

$\bar{\rho}_\star \simeq 3.9\times10^{5}\,\mathrm{g\,cm^{-3}}$, obtained from $\bar{\rho}_\star=3M_\star/(4\pi R_\star^3)$,
and assuming a rocky planet with Earth-like mean density $\bar{\rho}_{\rm p} \simeq 5.5\,\mathrm{g\,cm^{-3}}$, this expression yields a fluid Roche limit of

\begin{center}
$\displaystyle
a_{\rm Roche} \simeq 1.31\,R_\odot
\;\approx\; 6.1\times10^{-3}\,\mathrm{AU}.
$
\end{center}

Orbital radii interior to the fluid Roche limit are considered tidally unstable under the adopted fluid approximation. Accordingly, any nominal temperate regions within this limit are not interpreted as physically viable planetary orbits.

\begin{figure}[!t]
    \includegraphics[width=\linewidth]{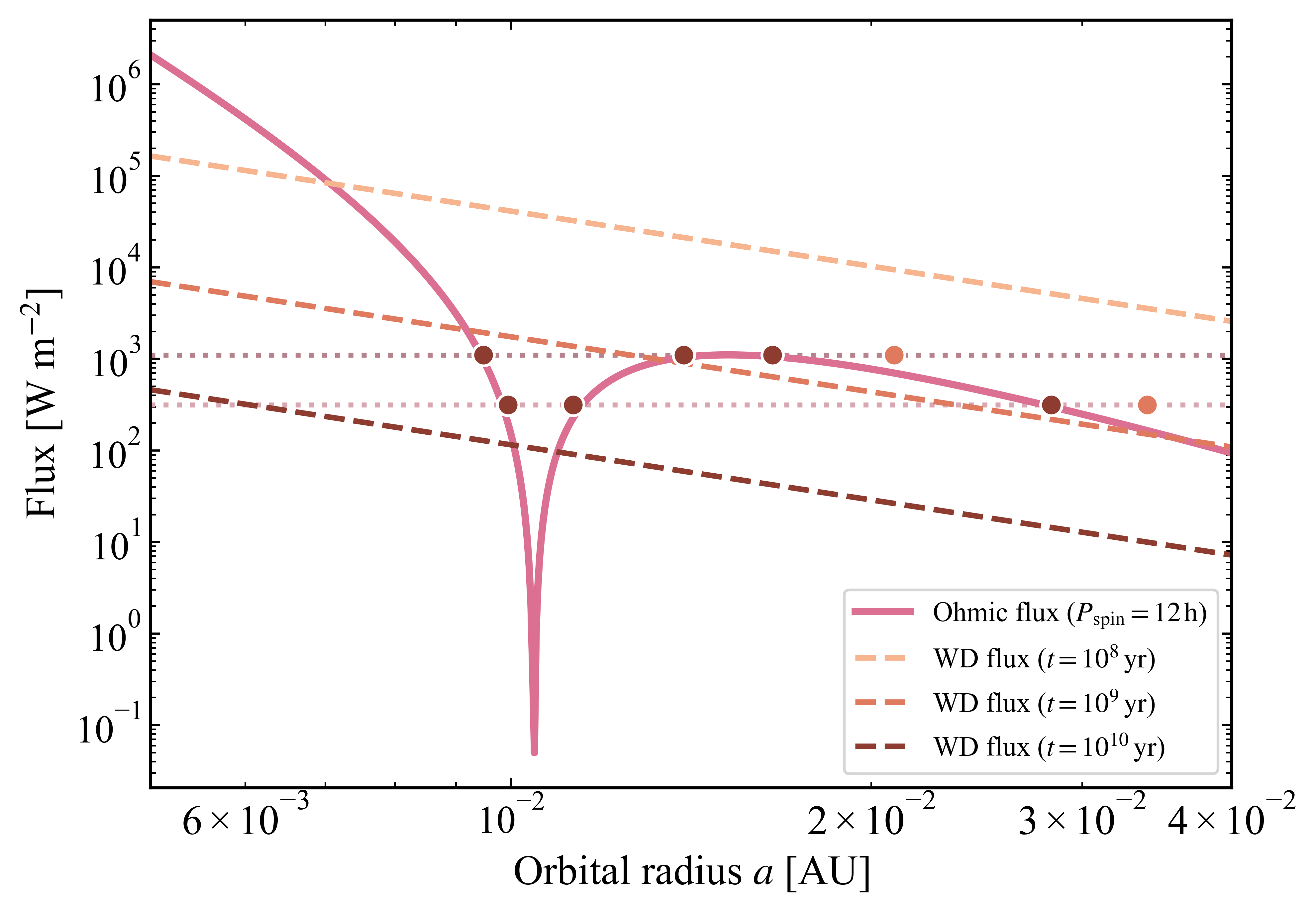}
    \caption{
    Stellar and induction heating fluxes as a function of orbital radius at three WD cooling ages.  
    Intersections between the total flux and the 273 and 373\,K reference levels (coloured markers) identify the evolving boundaries.}
    \label{fig:flux_comparison}
\end{figure}

Figure~\ref{fig:flux_comparison} provides a flux-based interpretation of the habitable zone structure shown in Figure~\ref{fig:multi_field_hz}.
For a given cooling time, the total surface flux received by a planet is the sum of stellar irradiation and induction heating.
Since a fixed habitable temperature corresponds to a fixed flux, identifying the habitable zone reduces to locating where the total flux intersects the flux levels for liquid water.

The plot compares three components of the planetary energy budget as a function of orbital radius. The solid pink curve represents the ohmic heating flux for the adopted high-field benchmark, with a surface dipole field strength of $B_\star = 100\,\mathrm{MG}$ and a fixed spin period of $P_{\rm spin}=12.0\,\mathrm{h}$. The three dashed curves show the stellar irradiation flux at cooling ages of $10^{8}$, $10^{9}$, and $10^{10}\,\mathrm{yr}$, with darker colors indicating progressively cooler white dwarfs. Horizontal dotted lines mark the flux levels associated with surface temperatures of $273\,\mathrm{K}$ and $373\,\mathrm{K}$.

At a given stellar age, the orbital radii at which the \emph{total} surface flux intersects these reference flux levels correspond to locations where a planet can maintain liquid water on its surface. The colored markers highlight these intersections, which therefore define the instantaneous habitable zone and illustrate how its extent evolves as the white dwarf cools and as induction heating contributes to the planetary energy budget.

At early times ($t \sim 10^{8}\,\mathrm{yr}$), stellar irradiation dominates and exceeds the habitable flux range throughout the plotted domain, producing no intersections. As the star cools ($t \sim 10^{9}$--$10^{10}\,\mathrm{yr}$), the stellar flux weakens and intersections begin to appear. The outer intersections correspond to the outer habitable zone seen in the temperature maps: a broad, slowly evolving region whose stability is enhanced by induction heating compensating for the declining stellar luminosity.

At very small orbital radii, an inner habitable zone emerges at late times. This occurs when irradiation weakens enough that induction heating alone can sustain habitable temperatures. Its narrow extent reflects the strong and non--monotonic dependence of the ohmic heating flux on the relative spin--orbit frequency $|\Omega_{\rm orb}-\Omega_{\rm spin}|$, which first increases and then declines with decreasing radius, allowing additional crossings with the habitable--flux levels on the inner side.

Overall, this flux--based view explains both the persistence of the outer habitable region and the appearance of a late--time inner window, and directly connects the contour structure in Fig.~\ref{fig:multi_field_hz} with the underlying energy budget of irradiated and induction--heated planets.

The detailed radial dependence of the induction heating flux and its connection to the relative spin--orbit motion are discussed further in Section~\ref{sec:habitability_dependence}.

\section{Results}
\label{sec:results}
\subsection{Consistency with Physical Limits}

Table~\ref{tab:phys_limits_example} summarizes the physical consistency checks for the fiducial system, including the sub-Alfv\'enic condition, Alfv\'en communication timescale, critical-resistance criterion, and Poynting-flux limit. All these conditions are satisfied for the adopted fiducial parameters, supporting the applicability of the closed-DC circuit approximation to this benchmark configuration.

\begin{deluxetable*}{lccc}
\tabletypesize{\scriptsize}
\tablewidth{0pt}
\tablecaption{Physical limit consistency checks.\label{tab:phys_limits_example}}
\tablehead{
  \colhead{Category} & \colhead{Quantity} & \colhead{Value} & \colhead{Status}
}
\startdata
\multirow{6}{*}{\textbf{Fiducial system}} & Surface field $B_\star$ [MG] & 100.00 & \nodata \\
& Spin period $P_{\rm spin}$ [h] & 12.00 & \nodata \\
& Orbital radius $a$ [AU] & 0.02 & \nodata \\
& Planet radius $R_{\rm p}$ [$R_\oplus$] & 1.00 & \nodata \\
& Planet mass $M_{\rm p}$ [$M_\oplus$] & 1.00 & \nodata \\
& Plasma density $n$ [cm$^{-3}$] & $1.00\times10^2$ & \nodata \\
\hline
\multirow{8}{*}{\textbf{Circuit resistances}} &
Planet total resistance $\mathcal{R}_{\rm p}$ [\,$\Omega$\,] & 1.50 & \nodata \\
& Atmospheric resistance $\mathcal{R}_{\rm atm}$ [\,$\Omega$\,] & 1.49887 & \nodata \\
& Ocean resistance $\mathcal{R}_{\rm ocean}$ [\,$\Omega$\,] & $5.00\times10^{-11}$ & \nodata \\
& Mantle resistance $\mathcal{R}_{\rm mantle}$ [\,$\Omega$\,] & $3.99\times10^{-4}$ & \nodata \\
& Core resistance $\mathcal{R}_{\rm core}$ [\,$\Omega$\,] & $6.00\times10^{-13}$ & \nodata \\
& Star resistance $\mathcal{R}_\star$ [\,$\Omega$\,] & 0.30 & \nodata \\
& Total resistance $\mathcal{R}_{\rm tot}$ [\,$\Omega$\,] & 1.80 & $\ge \mathcal{R}_{\rm crit}$ \\
& Critical resistance $\mathcal{R}_{\rm crit}$ [\,$\Omega$\,] & 0.43 & \checkmark \\
\hline
\multirow{4}{*}{\textbf{Velocities and Alfv\'enicity}} &
Orbital speed $v_{\rm orb}$ [m\,s$^{-1}$] & $1.64\times10^{5}$ & \nodata \\
& Relative speed $v_{\rm rel}$ [m\,s$^{-1}$] & $2.71\times10^{5}$ & \nodata \\
& Relativistic Alfv\'en speed $v_A$ [m\,s$^{-1}$] & $2.68\times10^{8}$ & \nodata \\
& Mach number $M_A=v_{\rm rel}/v_A$ & $1.01\times10^{-3}$ & \checkmark \\
\hline
\multirow{5}{*}{\textbf{Alfv\'en communication}} &
Alfv\'en round-trip time $\tau_A$ [s] & $2.23\times10^{1}$ & \nodata \\
& Crossing time $\tau_{\rm cross}$ [s] & $4.71\times10^{1}$ & \nodata \\
& Time ratio $\tau_A/\tau_{\rm cross}$ & 0.47 & \checkmark \\
& Critical density $n_{\rm crit}$ [cm$^{-3}$] & $1.84\times10^{3}$ & \checkmark \\
& Adopted communication condition & $\tau_A<\tau_{\rm cross}$ & \checkmark \\
\hline
\multirow{4}{*}{\textbf{EMF and power}} &
Induced EMF $\mathcal{E}$ [V] & $9.53\times10^{8}$ & \nodata \\
& Ohmic power in planetary $P_{\rm p}$ [W] & $4.20\times10^{17}$ & \nodata \\
& Total power $P_{\rm tot}$ [W] & $5.04\times10^{17}$ & $\le P_{\rm max}$ \\
& Maximum Poynting limit $P_{\rm max}$ [W] & $4.19\times10^{18}$ & \checkmark \\
\enddata

\tablecomments{The table presents consistency checks for the fiducial system. The relativistic Alfv\'en speed correction and communication conditions are detailed in Section~\ref{sec:physical_limits}.}

\end{deluxetable*}
\subsection{Habitability Dependence}
\label{sec:habitability_dependence}

\begin{figure}[!t]
    \centering
    \includegraphics[width=\linewidth]{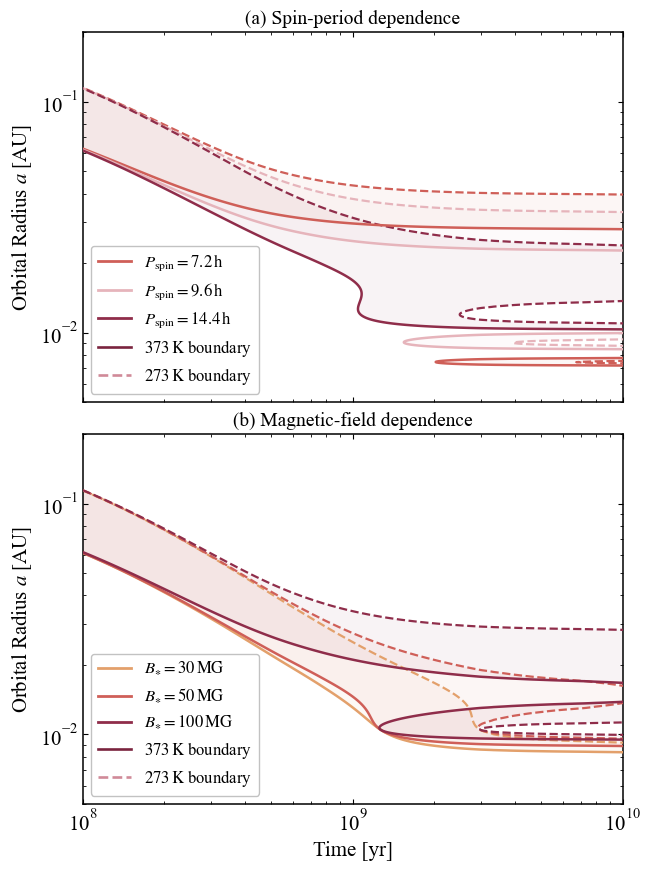}
    \caption{
    Dependence of temperate-zone evolution on stellar spin period and magnetic-field strength in the $f_{\rm dep}=1$ upper-limit calculation. 
    Panel (a) compares spin periods of
    $P_{\rm spin}=7.2$, $9.6$, and $14.4\,\mathrm{h}$
    at a fixed surface magnetic-field normalization of
    $B_\star=100\,\mathrm{MG}$.
    Panel (b) compares surface magnetic-field normalizations of
    $B_\star=30$, $50$, and $100\,\mathrm{MG}$
    at a fixed spin period of $P_{\rm spin}=12.0\,\mathrm{h}$.
    Solid curves denote the $373\,\mathrm{K}$ boundary and dashed curves denote the $273\,\mathrm{K}$ boundary.}
    \label{fig:hz_dependence}
\end{figure}

Figure~\ref{fig:hz_dependence} illustrates how the long-term evolution of temperate zones depends on the stellar spin period and magnetic-field strength. The solid and dashed curves correspond to the $373\,\mathrm{K}$ and $273\,\mathrm{K}$ boundaries, respectively, while the shaded regions indicate the resulting temperate zones during white dwarf cooling. Panel (a) fixes $B_\star=100\,\mathrm{MG}$ while varying $P_{\rm spin}$, whereas panel (b) fixes $P_{\rm spin}=12.0\,\mathrm{h}$ while varying $B_\star$, consistent with Table~\ref{tab:phys_limits_example}.

Figure~\ref{fig:hz_dependence}(a) shows the dependence on stellar spin period. Faster stellar rotation shifts the synchronous radius inward and increases the relative spin-orbit motion over most surviving planetary orbits. As a result, the upper-limit induction contribution remains larger at smaller orbital radii and later cooling ages. Systems with shorter spin periods therefore maintain broader temperate regions and exhibit significantly slower inward migration of the habitable zone under the efficient-deposition assumption. 
At late cooling ages, increasing $P_{\rm spin}$ shifts the outer temperate zone inward and the inner temperate zone outward, progressively narrowing the gap between them.

Figure~\ref{fig:hz_dependence}(b) shows the dependence on stellar magnetic-field strength. Stronger magnetic fields substantially enhance the induced electromotive force and increase the total ohmic dissipation throughout the system. Consequently, planets orbiting highly magnetized white dwarfs can sustain temperate conditions at larger orbital radii. 
For weaker magnetic fields, the outer temperate zone lies at smaller orbital radii and has a narrower radial extent at late cooling ages.

The comparison between panels (a) and (b) demonstrates that the stellar spin period and magnetic-field strength regulate habitability through physically distinct mechanisms. Variations in $P_{\rm spin}$ primarily modify the degree of spin-orbit asynchronism and the location of the synchronous radius, whereas variations in $B_{\ast}$ mainly control the overall amplitude of induction heating. Together, these parameters determine both the persistence and spatial extent of temperate environments around magnetic white dwarfs.
\subsection{Habitability Distribution}
\label{sec:relative_motion}

In the previous subsection we showed that ohmic heating depends sensitively on the relative motion between the planet and the stellar magnetic field. We now examine how this dependence translates into the cumulative duration over which a planet can maintain surface liquid water when the full white dwarf cooling history is taken into account.

\begin{figure}[!t]
    \centering
    \includegraphics[width=\linewidth]{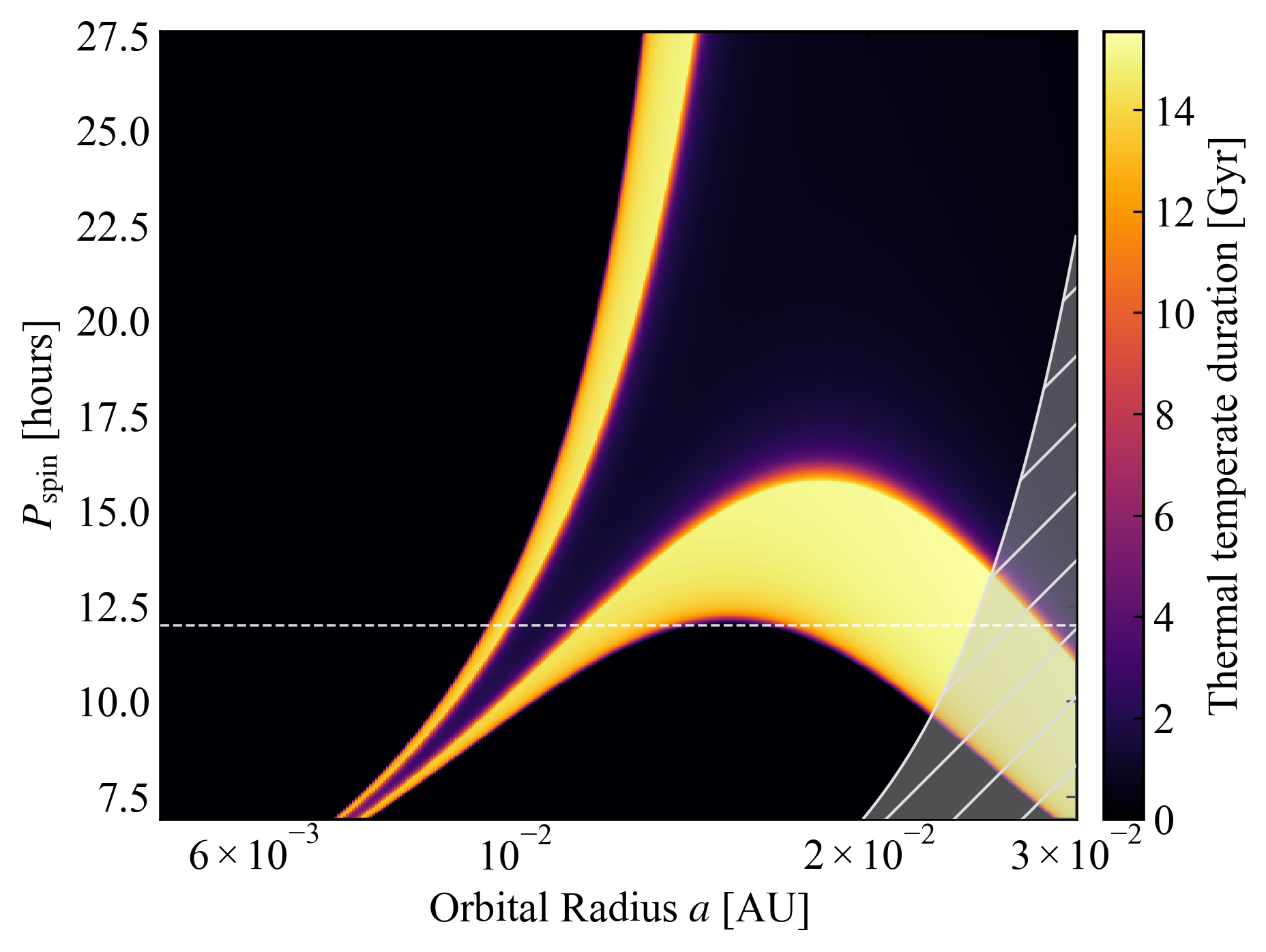}
    \caption{Two-dimensional map of cumulative temperate time as a function of orbital radius and white dwarf spin period. The stellar surface magnetic field is fixed at $B_\star=100\,\mathrm{MG}$. The shaded region denotes the parameter combinations that do not meet the communication constraint for a steady DC circuit, $\tau_A\geq\tau_{\rm cross}$.}
    \label{fig:wd_planet_habitability_map}
\end{figure}

Figure~\ref{fig:wd_planet_habitability_map} presents the cumulative duration over which a planet maintains liquid-water surface temperatures at fixed points in the $(a, P_{\rm spin})$ parameter space. For each grid point, we follow the full white dwarf cooling history and evaluate the planetary surface temperature at every snapshot using the net flux defined in Eq.~(\ref{eq:Fnet}). A snapshot is counted as habitable when the surface temperature lies between 273\,K and 373\,K, and summing over all snapshots yields the total habitable time shown by the colour scale.

The resulting distribution exhibits a pronounced wing-shaped structure, indicating that long-term habitability is confined to a narrow region where induction heating reaches an intermediate, moderate induction-heating regime. This pattern directly reflects the behaviour identified in Section~\ref{sec:habitability_dependence}: the induced electromotive force, and hence the ohmic dissipation, is controlled by the relative motion between the orbital frequency and the stellar spin frequency.

At small orbital radii, the stellar magnetic field stays strong and even modest relative motion can generate excessive ohmic heating. In this regime, habitable conditions are possible only when the white dwarf rotates sufficiently slowly, reducing the relative velocity and preventing runaway surface heating. This produces the left bright branch of the map, where large values of $P_{\rm spin}$ temper the induction power to levels compatible with surface liquid water.

At larger orbital radii, the magnetic coupling weakens rapidly and induction heating becomes insufficient unless the star rotates rapidly. This gives rise to the right bright branch, where short $P_{\rm spin}$ compensates for the reduced field strength and sustains adequate  planetary ohmic heating over gigayear timescales. Between these two branches lies a dark region corresponding to combinations of $(a, P_{\rm spin})$ that either overheat the surface or fail to provide enough deposited induction heating to offset the declining stellar irradiation.

Overall, this map demonstrates that the wing-shaped habitability pattern is a direct consequence of relative motion. Only specific combinations of orbital radius and stellar spin rate yield induction heating of the appropriate magnitude to balance stellar cooling and maintain habitable surface temperatures over gigayear timescales.

\subsection{Orbital and Surface Temperature Evolution}

To illustrate how specific planetary systems map onto the 2D structure shown in Figure~\ref{fig:wd_planet_habitability_map}, we next examine the coupled orbital and thermal evolution along three representative trajectories, with planets initially placed at $a_0 = 0.011, 0.015$, and $0.020$\,AU around a cooling white dwarf with fixed spin properties. We use a constant 100 MG surface field and $\mathcal{R}_{\rm tot}=1.80\,\Omega$; only $P_{\rm p}=I^2\mathcal{R}_{\rm p}$, with $\mathcal{R}_{\rm p}=1.50\,\Omega$, enters the planetary heat budget.

For the orbital-evolution calculation, we adopt an Earth-mass planet, $M_{\rm p}=1\,M_\oplus$, on a circular orbit. Following the unipolar-inductor framework \citep{Laine_2012,Lai_2012}, the magnetic torque acting on the planetary orbit is related to the total circuit power through $P_{\rm tot}=\Gamma_{\rm mag} (\Omega_{\rm spin}-\Omega_{\rm orb})$, giving

\begin{equation}
\label{eq:magnetic_torque}
\Gamma_{\rm mag}
=
-\frac{4R_{\rm p}^2B(a)^2a^2}
{\mathcal{R}_{\rm tot}}
(\Omega_{\rm orb}-\Omega_{\rm spin}).
\end{equation}

Here $\Omega_{\rm orb}=\sqrt{GM_\star/a^3}$.
For $M_{\rm p}\ll M_\star$, the orbital angular momentum
is $L_{\rm orb}=M_{\rm p}\sqrt{GM_\star a}$.
Using $dL_{\rm orb}/dt=\Gamma_{\rm mag}$,
the orbital migration equation becomes

\begin{equation}
\label{eq:orbital_migration}
\begin{aligned}
\frac{da}{dt}
&=
\frac{2\Gamma_{\rm mag}}
{M_{\rm p}\Omega_{\rm orb}a}\\
&=
-\frac{8R_{\rm p}^2B(a)^2a}
{M_{\rm p}\Omega_{\rm orb}\mathcal{R}_{\rm tot}}
(\Omega_{\rm orb}-\Omega_{\rm spin}).
\end{aligned}
\end{equation}

The magnetic torque drives outward orbital
migration when $\Omega_{\rm spin}>\Omega_{\rm orb}$.

At these close-in distances, the planets experience a persistent magnetic torque that drives a gradual radial migration \citep[e.g.,][]{Goldreich_1969, Li_1998, Lai_2012}. As the orbits drift, the evolving relative spin-orbit motion ($\Delta \Omega$) continuously modulates the induction heating power, a generic feature of star--planet magnetic interactions \citep{Laine_2012, Ahuir_2019}. Because the magnetic field strength scales as $a^{-3}$, outward migration reduces the local field and modifies both the irradiation and the ohmic dissipation along each trajectory.

\begin{figure}[!t]
    \centering
    \includegraphics[width=\linewidth]{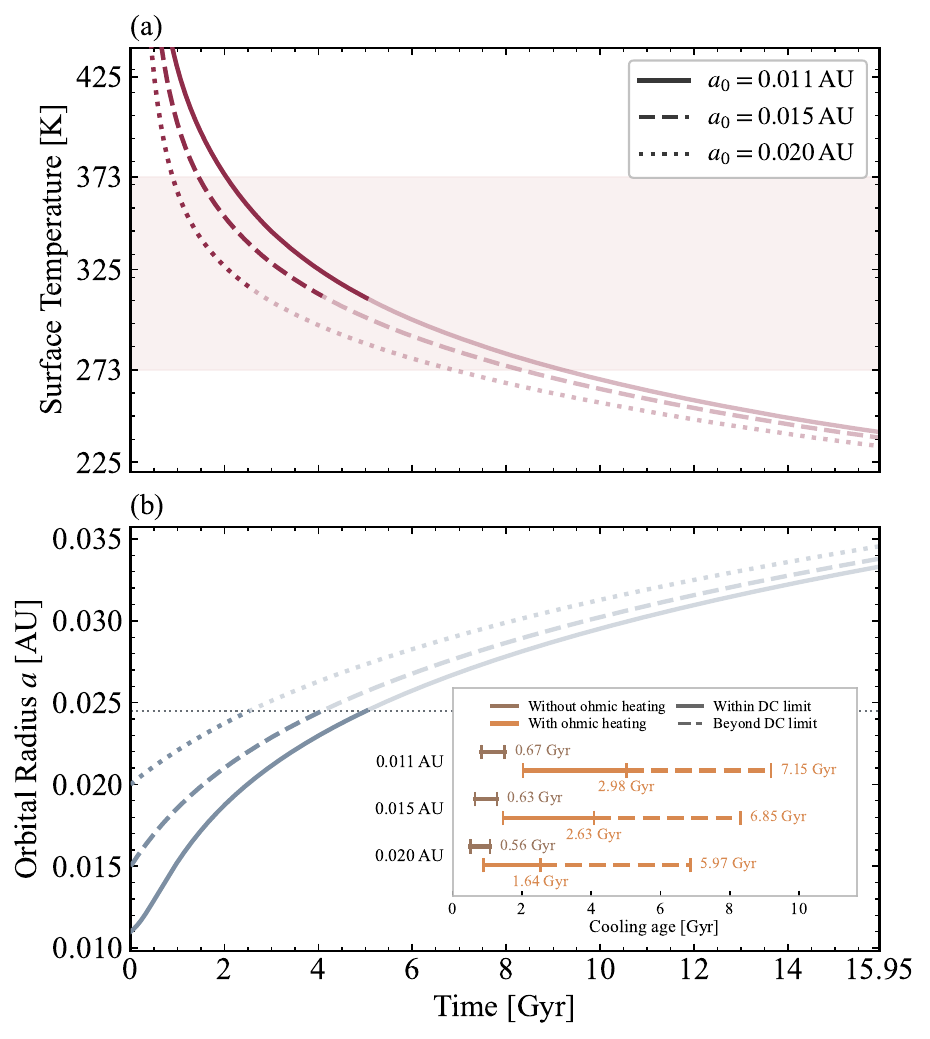}
    \caption{Coupled orbital and thermal evolution of three representative planets in the $f_{\rm dep}=1$ calculation.
    The adopted stellar magnetic field strength and spin period are $B_\star=100\,\mathrm{MG}$ and $P_{\rm spin}=12\,\mathrm{h}$, respectively.
    Dark red lines (upper panel) denote the planetary surface temperatures, while blue-gray lines (lower panel) track the corresponding orbital expansion. Different line styles (solid, dashed, and dotted) correspond to planets with initial orbital radii of $a_0 = 0.011, 0.015$, and $0.020$\,AU, respectively. The shaded band marks 273--373 K. Faded curves show formal DC extrapolations beyond the communication limit. The inset compares temperate intervals without and with ohmic heating along the same migrating orbits; solid and dashed segments lie within and beyond the communication limit, respectively. Labels give cumulative durations, and the horizontal axis shows cooling age.}
    \label{fig:trajectory}
\end{figure}

In principle, the same magnetic torques responsible for orbital migration also act back on the white dwarf spin \citep{Goldreich_1969,Lai_2012,Ahuir_2019}.
For each of the three planets, we compare the cumulative time spent in the 273--373\,K range with and without planetary ohmic heating along the same migrating trajectory. We include only those portions of each trajectory that satisfy the DC communication criterion of Equation~(18). Assuming efficient heat deposition ($f_{\rm dep}=1$) and a constant stellar footpoint resistance, we find that the cumulative temperate duration with ohmic heating is 2.9--4.4 times as long as with stellar irradiation alone. The dashed segments in the inset of Figure~\ref{fig:trajectory} show the additional temperate intervals obtained by extrapolating the DC model beyond the communication limit.

Along these gradually evolving orbits, we compute the planetary surface temperatures by combining stellar irradiation and induction heating, fully accounting for  the reduction of irradiating flux. As shown in Figure~\ref{fig:trajectory}, the initial orbital radius strongly influences the habitability timeline. Within the orbital segments satisfying the adopted communication criterion, including magnetic induction heating substantially increases the cumulative time spent in the 273--373 K temperate range for all three illustrative planets, compared with irradiation alone along the same migrating trajectories. This demonstrates that magnetic induction heating can extend temperate windows around white dwarfs even when orbital migration is included. Figure~{\ref{fig:trajectory}} also reports the full thermal intervals obtained without communication truncation.

\section{Discussion}
\label{sec:discussion}

\subsection{Comparison with Other Non-Radiative Sources}

Magnetic unipolar induction represents a physically distinct heating channel that complements tidal dissipation \citep{Heller_2013, Becker_2023} and delayed white dwarf cooling driven by $^{22}\mathrm{Ne}$ distillation \citep{Vanderburg_2025}. Unlike irradiation, induction depends primarily on the stellar field and spin--orbit asynchronism. In the efficient-deposition limit it can therefore supplement the surface energy budget after the white dwarf irradiation has declined.

\subsection{Stellar Footpoint Resistance}
\label{sec:rstar_sensitivity}

As introduced in Section~\ref{sec:unipolar_induction}, we evaluate the impact of a cooling-dependent stellar footpoint resistance $\mathcal{R}_{\star}(T_{\rm eff})$ to test the robustness of our constant-resistance baseline. By parameterizing the resistance to increase by a factor of $f_{\mathcal R}$ as the white dwarf cools to $3000\,\mathrm{K}$, we can quantify how a declining atmospheric electron fraction might suppress late-time induction heating.

\begin{figure}[!t]
    \centering
    \includegraphics[width=\linewidth]{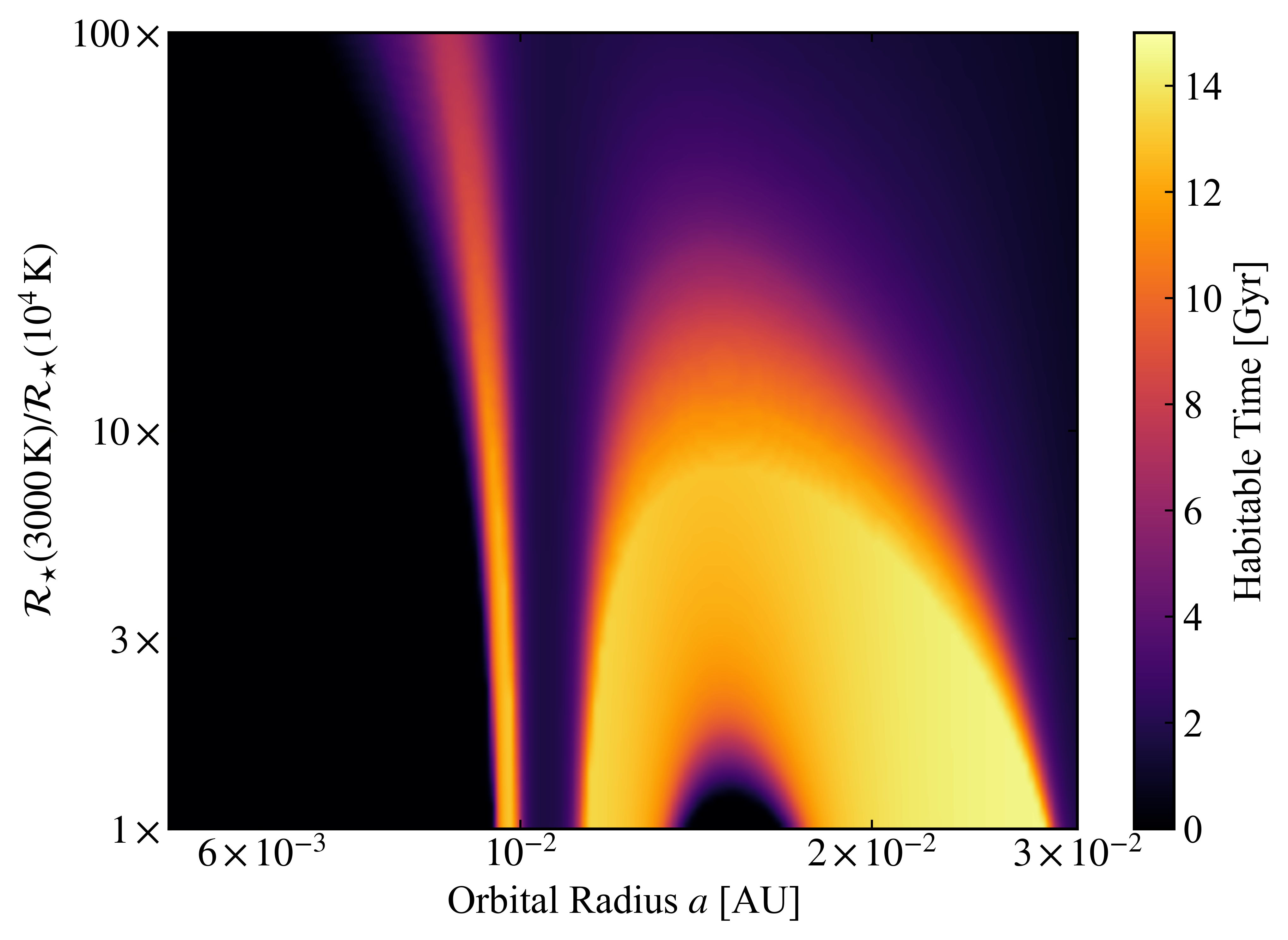}
    \caption{
Cumulative temperate duration as a function of orbital radius $a$ (horizontal axis, in AU) and the stellar footpoint resistance growth factor $f_{\mathcal R}=\mathcal{R}_{\star}(3000\,{\rm K})/\mathcal{R}_{\star}(10^4\,{\rm K})$ (vertical axis, logarithmic scale). The values $f_{\mathcal R}=1$, 3, 10, and 100 are marked on the vertical axis. The colorbar represents the cumulative duration (Gyr) for which the planetary surface temperature remains between 273 and 373\,K. The calculations adopt $B_\star=100\,\mathrm{MG}$, $P_{\rm spin}=12\,\mathrm{h}$, and $f_{\rm dep}=1$. Each map cell is evaluated over the full white dwarf cooling track.}
    \label{fig:rstar_sensitivity}
\end{figure}

These sensitivity calculations show how the longevity of the temperate-flux interval depends on the uncertain cooling evolution of the footpoint resistance. For the adopted $B_\star=100\,\mathrm{MG}$ and $P_{\rm spin}=12\,\mathrm{h}$ benchmark, the durations are evaluated over $f_{\mathcal R}=$1, 3, 10 and 100 and are shown in Figure~\ref{fig:rstar_sensitivity}.

Figure~\ref{fig:rstar_sensitivity} shows that the response is orbit dependent. Increasing $\mathcal{R}_{\star}$ lowers the planet-side power at fixed $(a,t)$, but can move an initially overheated close orbit into the temperate range. Modest growth ($f_{\mathcal R}\lesssim10$) still permits multi-Gyr temperate intervals in this parameterized calculation, whereas stronger growth substantially reduces or shifts them.

\FloatBarrier

\subsection{Observational Implications and Signatures}

To quantify these signatures, we estimate the observable temperature offset. For a planet where deposited induction heating contributes $F_{\rm eff}$, the effective temperature increases from $T_{\rm eq}$ to $T_{\rm eff}=(T_{\rm eq}^4+F_{\rm eff}/\sigma)^{1/4}$. Taking $F_{\rm eff}\sim\sigma(300\,\mathrm{K})^4$ and $T_{\rm eq}\sim200\,\mathrm{K}$ gives $T_{\rm eff}\sim320\,\mathrm{K}$. This is an upper-limit deposition estimate. The resulting thermal excess could enhance the secondary-eclipse depth in mid-infrared bands accessible to JWST.

In addition to thermal emission, induction heating also enhances the prospects for transmission spectroscopy by increasing the atmospheric scale height $H$. For an isothermal atmosphere, the scale height is given by $H = k_B T / (\mu m_H g)$, where $k_B$ is the Boltzmann constant, $T$ is the atmospheric temperature, $\mu$ is the mean molecular weight, and $g$ is the surface gravity. 

For a rocky planet with an Earth-like composition ($\mu \approx 29$) and surface gravity, an increase in temperature from $T_{\rm eq} \sim 200\,\mathrm{K}$ to $T_{\rm eff} \sim 320\,\mathrm{K}$ results in a 60\% expansion of the scale height (from $H \approx 5.8\,\mathrm{km}$ to $9.3\,\mathrm{km}$). Since the amplitude of transmission spectral features scales linearly with $H$, such atmospheric ohmic heating could significantly boost the signal-to-noise ratio for detecting atmospheric constituents around cooled white dwarfs. This effect is particularly critical for planets orbiting low-luminosity white dwarfs, where irradiation alone would result in extremely compact and difficult-to-detect atmospheres.

Magnetic star--planet interactions can also channel Poynting flux along the Alfv\'en wings toward the stellar footpoints, accelerating electrons that generate low-frequency radio emission via the electron cyclotron maser instability \citep[ECMI; e.g.,][]{Saur_2013, Vedantham_2020}. Crucially, the ECMI emission is powered by the Poynting flux carried along the Alfv\'en wings, whereas planetary ohmic dissipation ($P_{\rm p}$) represents a separate thermal-dissipation channel.

\subsection{Model Limitations and Uncertainties}

A primary structural assumption in our work is the one-dimensional series approximation for planetary resistance. This approach represents the current as flowing along equivalent radial paths connecting the day-side and night-side hemispheres, with each path traversing the layered body along a diameter-like direction. The actual current distribution is determined by the conductivity profile, the planetary geometry, and the locations and spatial extents of the current footprints. We therefore interpret the series resistance and layer-resolved dissipated powers as geometry-dependent effective quantities.

To quantify the geometric uncertainty introduced by this simplification, we solved $\nabla\cdot(\sigma\nabla\Phi)=0$ for the same layered planetary structure and compared the resulting three-dimensional ohmic dissipation with the one-dimensional estimate.

The computational domain extends to an outer shell radius of $R_{\rm out}=1.15R_{\rm p}$. The layered structure consists of a metallic core ($r<0.6R_{\rm p}$), a silicate mantle ($0.6R_{\rm p}<r<0.999R_{\rm p}$), a thin ocean layer ($0.999R_{\rm p}<r<R_{\rm p}$), and an atmospheric shell ($R_{\rm p}<r<R_{\rm out}$), with conductivities of $10^5$, $10^{-4}$, $2$, and $10^{-8}\,\mathrm{S\,m^{-1}}$, respectively. At the outer boundary, we impose the hemispheric current-flux condition $-\sigma\partial\Phi/\partial r=j_0\operatorname{sgn}(\cos\theta)$, where $j_0=1\,\mathrm{A\,m^{-2}}$ is an arbitrary normalization. The potential and normal current are continuous across layer interfaces, with regularity imposed at the center. We solve the axisymmetric conduction equation using a finite-volume method in spherical coordinates, with a layer-aligned radial grid and harmonic interface conductances. The resulting sparse linear system is solved using SciPy's \texttt{spsolve} after fixing the arbitrary potential reference. The effective resistance is calculated as $\mathcal{R}_{\rm p,3D}=P_{\rm ohm}/I^2$, where $P_{\rm ohm}=\int_V\sigma|\nabla\Phi|^2\,dV$ and $I=2\pi R_{\rm out}^2j_0$. Numerical convergence is verified through grid refinement.

We obtained $\mathcal{R}_{\rm p,3D}/\mathcal{R}_{\rm p,1D}=0.4$, corresponding to $\mathcal{R}_{\rm p,3D}=0.6\,\Omega$ for the fiducial parameters. With $\mathcal{R}_{\star}=0.3\,\Omega$, the true planetary ohmic power is therefore $\sim1.6$ times the one-dimensional value. We retain the one-dimensional prescription for the broad parameter survey and treat this factor as an order-unity normalization uncertainty.

Several simplifying assumptions in our model warrant further refinement. 
We represent planetary interiors with fixed-layered conductivities  and do not couple ohmic heating to melting, conductivity feedback, atmospheric transport, or escape. Although the three-dimensional benchmark described above accounts for radial current spreading, it does not resolve anisotropic conductivity or the detailed atmospheric current-closure and energy-transport processes required to determine $f_{\rm dep}$ self-consistently.
The fluid Roche limit also constrains the physical viability of close-in planetary orbits. For rocky planets with partially or extensively molten interiors, reduced material strength may allow greater tidal deformation, making the fluid Roche limit a useful conservative estimate of the tidal-disruption boundary. Consequently, the innermost temperature-based temperate regions may not correspond to surviving planets. The actual disruption threshold depends on the planetary internal structure, melt fraction, and material strength, which are not explicitly modeled in this work.

Additionally, our dipole model does not include the multipolar fields observed in some magnetic white dwarfs \citep{Bagnulo_2021}. We also treat the planet as an unmagnetized obstacle of radius $R_{\rm p}$. An intrinsic planetary field could enlarge the effective obstacle through the formation of a magnetosphere, alter the interaction efficiency through magnetic-field orientation and reconnection topology, and redirect field-aligned currents toward auroral or high-latitude regions \citep{Saur_2004,Saur_2013,Strugarek_2015,Strugarek_2016}. The plasma density and flux-tube slippage remain uncertain, and the long-term orbital evolution may also involve tides and planet--planet interactions \citep{Saur_2013,Veras_2016}. Despite these uncertainties, the magnitude of the effects identified here suggests that magnetic induction is a critical, yet overlooked driver of white dwarf habitability.

\section{Conclusions}
In this work, we investigated the role of magnetic unipolar induction in shaping the habitability of rocky planets orbiting white dwarfs. By coupling white dwarf cooling tracks with a unipolar--inductor model, we explored how induction-driven ohmic dissipation modifies the planetary surface energy budget and influences the long-term evolution of habitable zones.

Our main results can be summarized as follows.

First, in the efficient-deposition limit, induction heating can provide a long-lived non-radiative energy source for close-in rocky planets around magnetic white dwarfs. Unlike stellar irradiation, which decreases steadily as the white dwarf cools, the induction heating depends primarily on the stellar magnetic-field strength, orbital radius, spin--orbit asynchronism, and circuit resistances.

Second, the inclusion of induction heating significantly alters the conventional picture of white dwarf habitable zones. In addition to slowing the inward migration of the outer habitable zone, induction heating can also generate a narrow inner temperate region at late cooling ages, where liquid-water conditions are maintained primarily by  planetary ohmic dissipation rather than stellar irradiation in the adopted $f_{\rm dep}=1$ upper-limit calculation.

Third, the persistence and spatial extent of habitable conditions depend strongly on the stellar magnetic-field strength and spin period.
Stronger magnetic fields and greater spin--orbit asynchronism increase the induction power, thereby extending the duration of temperate conditions over gigayear timescales.

Fourth, the cumulative habitability map reveals a characteristic wing-shaped structure in the $(a, P_{\rm spin})$ parameter space. This structure arises naturally from the dependence of induction heating on the relative spin-orbit motion, demonstrating that long-term habitability requires a balance between excessive heating at small orbital radii and insufficient heating at large orbital separations.

Finally, we verified that the systems considered in this work remain within the physically self-consistent regime of the unipolar--inductor framework.

Overall, our results suggest that magnetic unipolar induction may represent an important and previously overlooked mechanism capable of reshaping white dwarf habitable zones and sustaining temperate conditions long after stellar irradiation alone would become insufficient.

Future work incorporating coupled interior evolution,  three-dimensional planetary current transport, conductivity feedback, atmospheric escape, and more realistic magnetic topologies will further clarify the role of magnetic star--planet interactions in the long-term evolution of planetary habitability around compact stars.

\begin{acknowledgments}
    {We thank the anonymous referee for constructive comments that helped improve the paper. } This work is supported by the National SKA Program of China (grant No. 2022SKA0120101), the National Natural Science Foundation of China (NSFC, No.12288102, 12403071), the science research grants from the China Manned Space Project (No. CMS-CSST-2021-B09, CMS-CSST-2021-B12, and CMS-CSST-2021-A10), International Centre of Supernovae (ICESUN), Yunnan Key Laboratory of Supernova Research (No. 202505AV340004), and the grants from The Macau Science and Technology Development Fund, and opening fund of State Key Laboratory of Lunar and Planetary Sciences (Macau University of Science and Technology) (Macau FDCT Grant No. SKL-LPS(MUST)-2021-2023). C.Y. has been supported by the National Science Foundation of China (grants 12373071). 
    X.W. acknowledges the use of ChatGPT (OpenAI) for assistance with English language editing, manuscript organization, and code debugging during the preparation of this work.
\end{acknowledgments}

\bibliography{Citation}{}
\bibliographystyle{aasjournalv7}

\end{document}